\documentclass[aps,pre,showkeys,groupedaddress,11pt]{revtex4}

\usepackage{graphicx,epic,eepic,epsfig,amsmath,latexsym,amssymb,verbatim}

\usepackage{theorem}
\newtheorem{definition}{Definition}
\newtheorem{proposition}[definition]{Proposition}
\newtheorem{example}[definition]{Example}
\newtheorem{lemma}[definition]{Lemma}

\def\squareforqed{\hbox{\rlap{$\sqcap$}$\sqcup$}}
\def\qed{\ifmmode\squareforqed\else{\unskip\nobreak\hfil
\penalty50\hskip1em\null\nobreak\hfil\squareforqed
\parfillskip=0pt\finalhyphendemerits=0\endgraf}\fi}
\def\endenv{\ifmmode\;\else{\unskip\nobreak\hfil
\penalty50\hskip1em\null\nobreak\hfil\;
\parfillskip=0pt\finalhyphendemerits=0\endgraf}\fi}
\newenvironment{proof}{\noindent \textbf{{Proof~} }}{\qed}

\mathchardef\ordinarycolon\mathcode`\:
\mathcode`\:=\string"8000
\def\vcentcolon{\mathrel{\mathop\ordinarycolon}}
\begingroup \catcode`\:=\active
  \lowercase{\endgroup
  \let :\vcentcolon
  }

\newcommand{\nc}{\newcommand}
\nc{\rnc}{\renewcommand}
\nc{\beq}{\begin{equation}}
\nc{\eeq}{{\end{equation}}}
\nc{\beqa}{\begin{eqnarray}}
\nc{\eeqa}{\end{eqnarray}}
\nc{\lbar}[1]{\overline{#1}}
\nc{\bra}[1]{\langle#1|}
\nc{\ket}[1]{|#1\rangle}
\nc{\ketbra}[2]{|#1\rangle\!\langle#2|}
\nc{\braket}[2]{\langle#1|#2\rangle}
\nc{\proj}[1]{| #1\rangle\!\langle #1 |}
\nc{\avg}[1]{\langle#1\rangle}
\rnc{\max}{\operatorname{max}}
\nc{\Rank}{\operatorname{Rank}}
\nc{\smfrac}[2]{\mbox{$\frac{#1}{#2}$}}
\nc{\tr}{\operatorname{Tr}}
\nc{\ox}{\otimes}
\nc{\dg}{\dagger}
\nc{\dn}{\downarrow}
\nc{\cA}{{\cal A}}
\nc{\cB}{{\cal B}}
\nc{\cC}{{\cal C}}
\nc{\cD}{{\cal D}}
\nc{\cE}{{\cal E}}
\nc{\cF}{{\cal F}}
\nc{\cG}{{\cal G}}
\nc{\cH}{{\cal H}}
\nc{\cI}{{\cal I}}
\nc{\cJ}{{\cal J}}
\nc{\cK}{{\cal K}}
\nc{\cL}{{\cal L}}
\nc{\cM}{{\cal M}}
\nc{\cN}{{\cal N}}
\nc{\cO}{{\cal O}}
\nc{\cP}{{\cal P}}
\nc{\cR}{{\cal R}}
\nc{\cS}{{\cal S}}
\nc{\cT}{{\cal T}}
\nc{\rU}{{\cal U}}
\nc{\cX}{{\cal X}}
\nc{\cZ}{{\cal Z}}
\nc{\csupp}{{\operatorname{csupp}}}
\nc{\qsupp}{{\operatorname{qsupp}}}
\nc{\var}{\operatorname{var}}
\nc{\rar}{\rightarrow}
\nc{\lrar}{\longrightarrow}
\nc{\polylog}{\operatorname{polylog}}
\nc{\supp}{\operatorname{supp}\,}
\nc{\1}{\openone}

\nc{\RR}{{{\mathbb R}}}
\nc{\CC}{{{\mathbb C}}}
\nc{\FF}{{{\mathbb F}}}
\nc{\NN}{{{\mathbb N}}}
\nc{\ZZ}{{{\mathbb Z}}}
\nc{\PP}{{{\mathbb P}}}
\nc{\QQ}{{{\mathbb Q}}}
\nc{\UU}{{{\mathbb U}}}
\nc{\WW}{{{\mathbb W}}}
\nc{\EE}{{{\mathbb E}}}
\nc{\id}{{\mathbb I}}

\nc{\ob}[1]{#1}

\begin{document}

\title{{\Large\sc All Inequalities for the Relative Entropy}}

\author{Ben Ibinson}
\email{ben.ibinson@bris.ac.uk}
\author{Noah Linden}
\email{n.linden@bristol.ac.uk}
\author{Andreas Winter}
\email{a.j.winter@bris.ac.uk} \affiliation{Department of
Mathematics, University of Bristol, Bristol BS8 1TW, United Kingdom}

\date{29th November 2005}

\begin{abstract}
  The relative entropy of two $n$-party quantum states is an
  important quantity exhibiting, for example, the extent to which
  the two states are different. The relative entropy of the states
  formed by reducing two $n$-party to a smaller number $m$ of
  parties is always less than or equal to the relative entropy of
  the two original $n$-party states. This is the monotonicity of
  relative entropy.

  Using techniques from convex geometry, we prove that
  monotonicity under restrictions is the only general inequality
  satisfied by relative entropies. In doing so we make a
  connection to secret sharing schemes with general access
  structures.

  A suprising outcome is that the structure of allowed relative
  entropy values of subsets of multiparty states is much simpler
  than the structure of allowed entropy values. And the structure
  of allowed relative entropy values (unlike that of entropies) is
  the same for classical probability distributions and quantum
  states.
\end{abstract}

\keywords{relative entropy, inequalities, cone, secret sharing.}

\maketitle

\section{Entropy and relative entropy}
\label{sec:intro}

Entropy inequalities play a central role in information
theory~\cite{Cover:Thomas}, classical or quantum. This is so
because practically all capacity theorems are formulated in terms
of entropy, and the same, albeit to a lesser degree, holds for many
monotones, of, for example, entanglement: e.g., the
\emph{entanglement of formation}~\cite{BDSW} or \emph{squashed
entanglement}~\cite{CW:04}. It may thus come as a surprise that
until recently~\cite{LW05} essentially the only inequality known
for the von Neumann entropies in a composite system is
\emph{strong subadditivity}
\begin{equation}
  \label{eq:ssa}
  S(\rho^{AB}) + S(\rho^{BC}) \geq S(\rho^{ABC}) + S(\rho^B),
\end{equation}
proved by Lieb and Ruskai~\cite{LR73}. We use the notation
$\rho^{ABC}$ for the density operator representing the state of
the system $ABC$, with the notation $\rho^{BC}=\tr_A\rho^{ABC}$
etc. for the reduced states.

The relative entropy of two states $\rho,\sigma$ (density
operators of trace $1$) is defined as
\begin{equation*}
  D(\rho\|\sigma) = \begin{cases}
                      \tr \rho(\log\rho-\log\sigma) & \text{ if }
                                                      \supp\rho\subset\supp\sigma, \\
                      +\infty                       & \text{ otherwise},
                    \end{cases}
\end{equation*}
where $\supp\rho$ is the supporting subspace of the density
operator $\rho$. Note that in this paper, log always denotes the
logarithm to base 2. Like von Neumann entropy, the relative
entropy is used extensively in quantum information and
entanglement theory to obtain capacity-like quantities and
monotones. The most prominent example may be the \emph{relative
entropy of entanglement}~\cite{VedPlen,VPRK}. Many other
applications of the relative entropy are illustrated in the
review~\cite{Vedral}.

In this paper we study the universal relations between the
relative entropies in a composite system and for general pairs of
states. For the most part we shall restrict ourselves to finite
dimensional spaces.

What are the known inequalities? First of all, the relative
entropy is always nonnegative, and indeed $0$ iff $\rho=\sigma$
(see the recent survey by Petz~\cite{Petz}). The most important,
and indeed only known inequality, for the relative entropy is the
\emph{monotonicity},
\begin{equation}
  \label{eq:mono}
  D(\rho^{AB}\|\sigma^{AB}) \geq D(\rho^A\|\sigma^A)
\end{equation}
for a bipartite system $AB$. This relation can be derived from
strong subbadditivity, eq.~(\ref{eq:ssa}), as was shown by
Lindblad~\cite{Lindblad} in the finite dimensional case;
Uhlmann~\cite{Uhlmann} later showed it in generality. To
illustrate the connection, strong subadditivity can be easily
derived from eq.~(\ref{eq:mono}). Note that we can identify the
following relative entropy quantity with \emph{quantum mutual
information}~\cite{CA:97}:
\begin{align*}
  D(\rho^{AB} \| \rho^A \otimes \rho^B)
     &= \tr{\big(\rho^{AB} \log\rho^{AB}\big)}
           - \tr{\big(\rho^{AB} \log (\rho^A \otimes \rho^B)\big)}\\
     &=-S(\rho^{AB})- \tr{\big(\rho^{AB} \log \rho^A\big)}
           - \tr{\big(\rho^{AB} \log \rho^B\big)}                 \\
     &=-S(\rho^{AB})+S(\rho^A)+S(\rho^B)                          \\
     &=I(A:B).
\end{align*}
Hence we can recover strong subadditivity from the monotonicity
relation
\begin{equation*}
D(\rho^{ABC}\|\rho^{AB} \otimes \rho^C)  \geq D(\rho^{BC}\|\rho^B
\otimes \rho^C),
\end{equation*}
as follows:
\begin{align*}
  0 &\leq D(\rho^{ABC}\|\rho^{AB} \otimes \rho^C)
               - D(\rho^{BC}\|\rho^B\otimes \rho^C) \\
    &=I(AB:C) - I(B:C)                              \\
    &=S(\rho^{AB})+S(\rho^C)-S(\rho^{ABC})- S(\rho^B)-S(\rho^C)+S(\rho^{BC})\\
    &=S(\rho^{AB})+S(\rho^{BC}) - S(\rho^{ABC})-S(\rho^B).
\end{align*}
Before returning to relative entropy we make a few further
observations about entropy. For an $n$-party system, there are
$2^n-1$ non-trivial reduced states, with their entropies, so we
can associate with each state a vector of $2^n-1$ real
coordinates. Pippenger~\cite{Pippenger}, following the programme
of Yeung and Zhang in the classical case~\cite{Yeung}, showed
that, after going to the topological closure, the set of all
entropy vectors is a convex cone. Hence it must be describable by
linear (entropy) inequalities, like strong subadditivity, and one
can ask if the entropy cone coincides with the cone defined by the
"known" inequalities (strong subadditivity in the quantum case,
additionally positivity of conditional entropy classically). This
is indeed the case for $n\leq 3$: the classical result is due to
Yeung and Zhang~\cite{Yeung}, the quantum case by
Pippenger~\cite{Pippenger}. Yeung and Zhang~\cite{YZ} have however
found a new, "non-Shannon type" inequality for $n=4$ classical
parties, and Linden and Winter~\cite{LW05} found a new so-called
constrained inequality for $n=4$ quantum parties, providing
evidence that to describe the entropy cones of four and more
parties one needs new inequalities, too.

In~\cite{LW05} Linden and Winter describe how the putative vector
of entropies,
\begin{equation}
\label{eq:ray}
 [S_A,S_B,\ldots,S_{ABCD}] = \lambda [3,3,2,2,4,3,3,3,3,4,4,4,3,3,2],
\end{equation}
for $\lambda \geq 0$, satisfies strong subadditivity for all subsets
of parties $ABCD$, but is nonetheless not achievable by any quantum
state $\rho$ [i.e. there is no quantum state $\rho$ such that
$S_A=S_A(\rho_A), S_B=S_B(\rho_B)$ etc. achieving the values in
eq.~(\ref{eq:ray})]. Here we ask (and answer in the affirmative) the
question of whether any vector
\begin{equation*}
[D_A,D_B,\ldots,D_{ABCD}]
\end{equation*}
in which the numbers $D_A,\ldots,D_{ABCD}$ satisfy the constraints
of monotonicity for all subgroups may be realised as the relative
entropy of pairs of states [i.e. for any such vector we show that
there are states $\rho$ and $\sigma$ such that $D_A =
D(\rho_A\|\sigma_A)$ etc.]

In this paper we prove the result that for relative entropy,
monotonicity is necessary and sufficient to describe the complete
set of realisable relative entropy vectors. This is a surprising
discovery as relative entropy is a seemingly more complex
functional than entropy. However strong subadditivity is
sufficient to define all possible relative entropy vectors (as
monotonicity is derived from it) whereas it cannot encapsulate
normal von Neumann entropy. Our approach is as follows: we show
first, by adapting the Yeung-Pippenger techniques, that the
topological closure of the set of all relative entropy vectors is
a convex cone (section~\ref{sec:relent-cone}). Then we study the
extremal rays of the Lindblad-Uhlmann cone defined by
monotonicity, in section~\ref{sec:LU-cone}: they correspond
one-to-one to so-called up-sets in $2^{[n]}$. It remains to prove
that every one of the rays is indeed populated by relative entropy
vectors, which we do in section~\ref{sec:equality}. It turns out
that the construction to show this depends heavily on secret
sharing schemes, which we explain in section~\ref{sec:equality},
to make the paper self-contained, followed by an instructive
example in section~\ref{sec:thres}, after which we conclude.

\section{The cone of relative entropy vectors}
\label{sec:relent-cone} Define the set $\Lambda_n^* \subset
\RR_{\geq 0}^{2^n -1}$ of vectors $\mathbf{v} = \bigl( v_{\cal S}
\bigr)_{\emptyset\neq{\cal S}\subset[n]}$ , with $[n] =
\{1,2,\ldots,n\}$: $\mathbf{v}\in \Lambda_n^*$ iff there exist
quantum states of $n$-parties $\rho,\sigma$ such that
$D(\rho^{\cal S}\|\sigma^{\cal S})=v_{\cal S}$ for every non empty
subset $\cal S$. Observe that there are $2^n-1$ nonempty subsets
${\cal S}$, which label the coordinates of $\RR^{2^n -1}$ in some
fixed way.

\begin{lemma}
  \label{lemma:cone}
  The topological closure $\overline{\Lambda_n^*}$ of $\Lambda_n^*$ is
  a convex cone. To be precise, it is enough to show that~\cite{Pippenger}:
  \begin{enumerate}
    \item (Additivity) for $\mathbf{v},\mathbf{w}\in \Lambda_n^*$,
      $\mathbf{v}+\mathbf{w}\in\Lambda_n^*$;

    \item (Approximate diluability) for all $\delta>0$ there exists $\epsilon>0$ such that
      for all $\mathbf{v}\in\Lambda_n^*$ and $0 \leq \lambda \leq \epsilon$
      there is $\mathbf{w}\in\Lambda_n^*$ with $\| \lambda \mathbf{v} - \mathbf{w} \| \leq
      \delta$.
  \end{enumerate}
  (We use the sup norm in the proof below, but since all norms in finite dimensions
  are equivalent, the exact choice of the norm is irrelevant.)
\end{lemma}
\begin{proof}
  Consider the following states $\rho , \rho', \sigma$ and
  $\sigma'$ where the prime indicates that the corresponding state lives on a system
  different from the unprimed states. Let us define $\mathbf{v}$ and $\mathbf{v}'$ as the
  relative entropy vectors generated from taking entropy values of
  $D(\rho^{\cal S}\|\sigma^{\cal S})$ and $D(\rho^{'\cal S}\|\sigma^{'\cal
  S})$ respectively. Consider states $\widetilde{\rho}=\rho \otimes
  \rho'$ and $\widetilde{\sigma}=\sigma \otimes
  \sigma'$. To prove the first part of the Lemma, we show $\widetilde{\mathbf{v}} =
  \mathbf{v}'+\mathbf{v}$ for the relative entropy vector
  $\widetilde{\mathbf{v}}$ of $\widetilde{\rho}$; in detail,
  for every ${\cal S} \subset [n]$,
  \begin{equation*}
    D(\widetilde{\rho}^{\cal S} \| \widetilde{\sigma}^{\cal S})=
    D(\rho^{\cal S} \otimes \rho'^{\cal S} \| \sigma^{\cal S} \otimes \sigma'^{\cal S} ) =
    D(\rho^{\cal S} \| \sigma^{\cal S}) + D(\rho'^{\cal S} \| \sigma'^{\cal S}).
  \end{equation*}
  Then,
  \begin{align*}
  D(\widetilde{\rho}^{\cal S} \| \widetilde{\sigma}^{\cal S})
                     =& -S(\widetilde{\rho}^{\cal S})
                        -\tr(\widetilde{\rho}^{\cal S} \log \widetilde{\sigma}^{\cal S}) \\
                     =& -S(\rho^{\cal S}) - S(\rho'^{\cal S}) - \tr\big(\rho^{\cal S} \ox
                     \rho'^{\cal S} \log \big(\sigma^{\cal S} \ox \sigma'^{\cal S}\big)\big).
  \end{align*}
  We use the fact that
  $\log(\sigma\otimes\sigma')=(\log \sigma)\otimes \1 + \1\otimes(\log\sigma')$.
  Therefore,
  \begin{align*}
    D(\widetilde{\rho}^{\cal S} \| \widetilde{\sigma}^{\cal S})
                    =& -S(\rho^{\cal S}) - S(\rho'^{\cal S})
                       -\tr(\rho^{\cal S} \log \sigma^{\cal S}) - \tr(\rho'^{\cal S} \log
                                  \sigma'^{\cal S})\\
                    =& D(\rho^{\cal S}\|\sigma^{\cal S}) + D(\rho'^{\cal S}\|\sigma'^{\cal S}).
  \end{align*}
  Therefore we can always construct a state that will give a vector
  in $\Lambda_n^*$ and is the sum of $\mathbf{v}$ and
  $\mathbf{v}'$.

  To prove the second part, choose $\epsilon$ such that $\epsilon
  \leq 1/2$ and $H_2(\epsilon) \leq \delta$ where $H_2(\epsilon)$ is
  the binary entropy of $\epsilon$,
  $$H_2(\epsilon)=-\epsilon \log \epsilon -(1-\epsilon)\log
  (1-\epsilon).$$
  Note that we can always
  choose a value of $\epsilon$ which satisfies these conditions for
  any $\delta$. Let $\mathbf{v}$ be the relative entropy vector created by states $\rho,
  \sigma$. Consider the following states,
  $\widehat{\rho}=\lambda \rho+(1-\lambda) \sigma$ and $\widehat{\sigma}=\sigma$
  with the entropy vector $\mathbf{w}$ created by states $\widehat{\rho},
  \widehat{\sigma}$. Consider the following quantity that leads to the entropy vector
  $\mathbf{w}$:
  \begin{align}
  \label{eq:fol}
  D(\widehat{\rho}^{\cal S} \| \widehat{\sigma}^{\cal S})
                  =& D\big(\lambda \rho^{\cal S} + (1-\lambda) \sigma^{\cal S} \|
                                                            \sigma^{\cal S}\big)\nonumber\\
                  =& -S\big(\lambda \rho^{\cal S} + (1-\lambda) \sigma^{\cal S}\big)
                     -\tr[\big(\lambda \rho^{\cal S} + (1-\lambda) \sigma^{\cal S}\big)\log
                                                             \sigma^{\cal S}]\nonumber\\
                  =&-S\big(\lambda \rho^{\cal S} + (1-\lambda) \sigma^{\cal S}\big) - \lambda
                  \tr(\rho^{\cal S}
                  \log \sigma^{\cal S}) - (1-\lambda)\tr(\sigma^{\cal S} \log
                  \sigma^{\cal S}).
  \end{align}
  We now make use of the following inequality, see for example~\cite{Neison:Chang}.
  \begin{equation*}
    \sum_i p_i S(\rho_i) \leq S\bigg(\sum_i p_i \rho_i\bigg) \leq
    H(p_i) + \sum_i p_i S(\rho_i),
  \end{equation*}
  which here specialises to
  \begin{equation*}
    \lambda S(\rho^{\cal S}) + (1-\lambda) S(\sigma^{\cal S})  \leq S\big(\lambda \rho^{\cal S} + (1-\lambda) \sigma^{\cal S}\big)  \leq H_2(\lambda) + \lambda
    S(\rho^{\cal S})+ (1-\lambda) S(\sigma^{\cal S}).
  \end{equation*}
  Hence we can  define a quantity $\alpha$ such that $0 \leq \alpha \leq
    H_2(\lambda) \leq H_2(\epsilon)\leq \delta$
  \begin{equation*}
     S\big(\lambda \rho^{\cal S} + (1-\lambda) \sigma^{\cal S}\big) = \lambda
    S(\rho^{\cal S})+ (1-\lambda) S(\sigma^{\cal S})+\alpha.
  \end{equation*}
  Therefore, eq.~(\ref{eq:fol}) reads,
  \begin{align*}
    D(\widehat{\rho}^{\cal S} \| \widehat{\sigma}^{\cal S})
                   =&-S\big(\lambda \rho^{\cal S} + (1-\lambda) \sigma^{\cal S}\big) - \lambda
                                                                             \tr(\rho^{\cal S}
                  \log \sigma^{\cal S}) - (1-\lambda)\tr(\sigma^{\cal S} \log \sigma^{\cal S})\\
                   =&-\lambda S(\rho^{\cal S})- (1-\lambda) S(\sigma^{\cal S}) - \alpha- \lambda
                   \tr(\rho^{\cal S}
                  \log \sigma^{\cal S}) - (1-\lambda)\tr(\sigma^{\cal S} \log \sigma^{\cal S})\\
                   =&  \lambda D(\rho^{\cal S}\|\sigma^{\cal S}) + (1-\lambda)
                  D(\sigma^{\cal S}\|\sigma^{\cal S})- \alpha\\
                   =& \lambda D(\rho^{\cal S}\|\sigma^{\cal S})-\alpha.
  \end{align*}
  Thus for our given vector $\mathbf{v}$ [the vector made from
  the relative entropies $D(\rho^{\cal S}\|\sigma^{\cal S})$], we
  have found a $\mathbf{w}$ [the vector of the relative
  entropies $D(\widehat{\rho}^{\cal S} \| \widehat{\sigma}^{\cal S})$]
  such that for all $\delta>0$ (where $H(\epsilon) \leq \delta$),
  \begin{equation*}
    \| \lambda \mathbf{v}-\mathbf{w} \| = \alpha \leq \delta
  \end{equation*}
  for all $\lambda \leq \epsilon$ (where $H(\epsilon) \leq \delta$).
  This completes the proof.
\end{proof}

\section{The Lindblad-Uhlmann cone}
\label{sec:LU-cone} Define the convex cone $\Lambda_n \subset
\RR_{\geq 0}^{2^n -1}$: all vectors $\mathbf{v}$ satisfying the
following inequalities, for all $[n] \supset {\cal S} \supset
{\cal S}' \neq \emptyset$:
\begin{align}
  \label{eq:mon1}
  v_{\cal S} &\geq v_{{\cal S}'},\\
  \label{eq:mon2}
  v_{\cal S} &\geq 0.
\end{align}
This defines the cone of all vectors that obey the only known
inequality between relative entropies of subsystems, the
Lindblad-Uhlmann monotonicity relation (which implies
non-negativity).

\begin{proposition}
  \label{prop:Lambda-extremal}
  The extremal rays of $\Lambda_n$ are spanned by vectors $\mathbf{u}$ of the form
  \begin{equation*}
    u_{\cal S} = \begin{cases}
                   1 & \text{ if }{\cal S}\in{\bf U},  \\
                   0 & \text{ if }{\cal S}\not\in{\bf U},
                 \end{cases}
  \end{equation*}
  for a set family $\emptyset \neq {\bf U} \subset 2^{[n]}$ and
  $\emptyset \notin {\bf U}$
  with the property that for all ${\cal S}\in{\bf U}$ and
  ${\cal S}' \supset {\cal S}$, ${\cal S}'\in{\bf U}$.
  (Such a set family is called an \emph{up-set}.)

  Conversely, every up-set ${\bf U}$, by the above
  assignment, defines a vector $\mathbf{u} \in \Lambda_n$ spanning an extremal ray.
\end{proposition}
\begin{proof}
Every extremal ray $R$ of $\Lambda_n$ is spanned by a vector
$v\in\Lambda_n$, such that $R=\RR_{\geq 0}\,\mathbf{v}$. It has the
property that if $\lambda\mathbf{a}+\mu\mathbf{b} \in R$ for
$\lambda,\mu > 0$ and $\mathbf{a},\mathbf{b} \in \Lambda_n$, then
$\mathbf{a},\mathbf{b} \in R$. With this every point in the cone is
a positive linear combination of elements from extremal rays. In
geometric terms, $R$ is an edge of the cone $\Lambda_n$
\cite{grunbaum}. It is a standard result from convex geometry (see
\cite{grunbaum}) that an extremal ray is specified by requiring that
sufficiently many of the defining inequalities are satisfied with
equality, in the sense that the solution space of these equations is
one-dimensional. (Of course, in addition the remaining inequalities
must hold.)

In the present case, there are only two, very simple, types of
inequalities. For a spanning vector $\mathbf{v}$ of an extremal
ray $R$, the equations (i.e., inequalities satisfied with
equality) take one of the following two forms: for $\mathcal{A}
\subset \mathcal{B}$, $\mathcal{C} \subset [n]$,
\begin{align}
  \label{eq:ineq}
  v_{\mathcal{A}} &= v_{\mathcal{B}},\\
  \label{eq:ineq2}
  v_{\mathcal{C}} &= 0.
\end{align}

How can it be that $\mathbf{v}$ is specified by a set of such
equations up to a scalar multiple? Since the equations only demand
that an entry of $\mathbf{v}$ is $0$ or that two entries are
equal, it must be such that there exists a subset $\mathbf{U}
\subset 2^{[n]}$ such that for all $\mathcal{A},\mathcal{B} \in
\mathbf{U}$, the corresponding entries of $\mathbf{v}$ are equal,
$v_{\mathcal{A}}=v_{\mathcal{B}}=v$, while for
$\mathcal{C}\not\in\mathbf{U}$, it holds that $v_{\mathcal{C}}=0$.
Now, to satisfy all the monotonicity inequalities, $\mathbf{U}$
must be an up-set. (We note that $\mathbf{v}\neq\mathbf{0}$ to
span a ray, hence $v\neq 0$.)

Thus, $\mathbf{v} = v \mathbf{u}$ for the vector $\mathbf{u}$
constructed from the up-set $\mathbf{U}$ in the statement of the
Proposition. This shows that every extremal ray is determined by
an up-set.

For the other direction, we first observe that $\mathbf{u}$
constructed from an arbitrary up-set $\mathbf{U}$ as stated
satisfies all the inequalities. Furthermore, it is clear that many
inequalities will be saturated. To show that $R=\RR_{\geq
0}\,\mathbf{u}$ is extremal, we only need to find a set of $2^n -
2$ linearly independent equations of the form~(\ref{eq:ineq})
and~(\ref{eq:ineq2}) that are satisfied. This is given by
\begin{alignat*}{2}
  v_{\mathcal{A}} &= v_{[n]} &\quad&\mbox{for $[n]\neq \mathcal{A} \in \mathbf{U}$}, \\
  v_{\mathcal{B}} &= 0       & &\mbox{for $\mathcal{B}\not\in \mathbf{U}$}.
\end{alignat*}
Indeed, these equations leave only the freedom to choose
$v_{[n]}$, and then all entries of $\mathbf{v}$ are determined.
This concludes the proof that every up-set determines an extremal
ray.
\end{proof}
\begin{example}
The following table shows all the extremal rays and hence all
possible up-sets for three parties up to permutations of parties.

\begin{center}
\begin{tabular}[c]{|c|c|c|c|c|c|c|c|}
    \hline
    \ &$v_A$ & $v_B$ & $v_C$ & $v_{AB}$ & $v_{AC}$ & $v_{BC}$ & $v_{ABC}$\\
    \hline
    \hline
    \ Ray 1&0    &   0  &   0  &   0   &   0   &   0   &    1\\
    \ Ray 2&0    &   0  &   0  &   1   &   0   &   0   &    1\\
    \ Ray 3&0    &   0  &   0  &   1   &   1   &   0   &    1\\
    \ Ray 4&0    &   0  &   0  &   1   &   1   &   1   &    1\\
    \ Ray 5&1    &   0  &   0  &   1   &   1   &   0   &    1\\
    \ Ray 6&1    &   0  &   0  &   1   &   1   &   1   &    1\\
    \ Ray 7&1    &   1  &   0  &   1   &   1   &   1   &    1\\
    \ Ray 8&1    &   1  &   1  &   1   &   1   &   1   &    1\\
    \hline
\end{tabular}
\end{center}
These up-sets are also represented in graphical form in Fig.~1.

\begin{figure}
\begin{center}
\includegraphics[scale=0.42,clip]{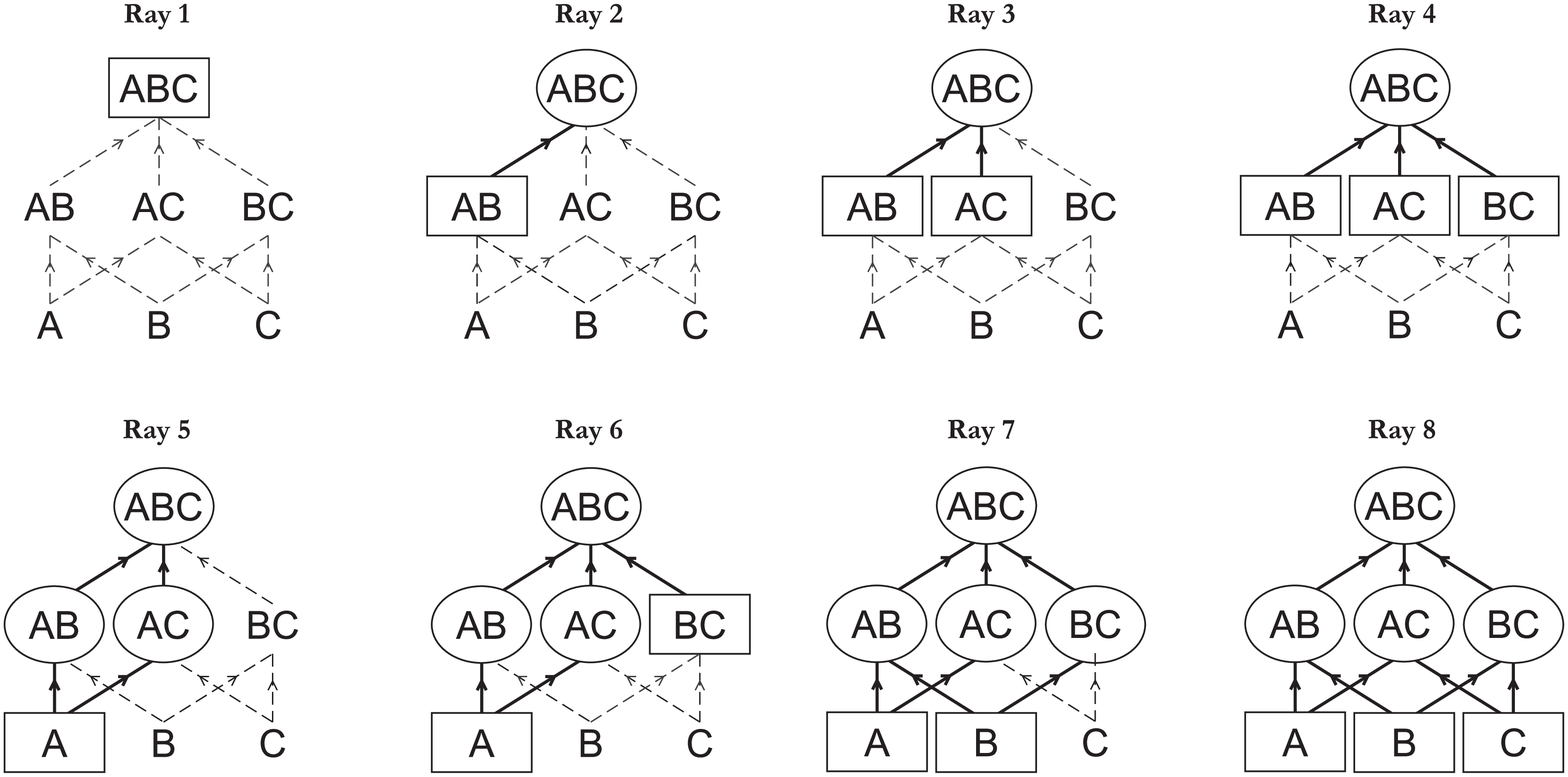}
\end{center}
\begin{caption}
{All possible `up-sets' for three parties, up to permutations.
Broken arrows indicate which sets have the corresponding element
as a subset, i.e broken arrow implies is subset of. Every element
that is inside a box or circle is defined as having relative
entropy 1. A box indicates that we have chosen the set to have
relative entropy 1, where as a circle indicates the set is forced
to have relative entropy 1 as one of its subsets also has relative
entropy 1. This `forcing' of relative entropy via one of the
subsets is represented as a black arrow.}
\end{caption}
\end{figure}
\end{example}

Note that every extremal ray of the relative entropy cone is very
well structured and can be defined precisely with up-sets. The
standard entropy cone however shows no such structure and its
extremal rays, although realised by highly structured states, show
far less structure in the actually entropy values of the extremal
rays (see~\cite{Pippenger,Magnificent:7}).

\section{$\mathbf{ \overline{\Lambda_n^*} = \Lambda_n }$}
\label{sec:equality} Clearly $\Lambda_n^* \subset \Lambda_n$ since
all actual states obey the Lindblad-Uhlmann monotonicity
inequalities~(\ref{eq:mon1}) and (\ref{eq:mon2}). Since $\Lambda_n$ is
closed, we thus get $\overline{\Lambda_n^*} \subset \Lambda_n$.

In this section we will show the opposite inclusion,
$\overline{\Lambda_n^*} \supset \Lambda_n$, thus showing
equality between the relative entropy cone and the Lindblad-Uhlmann cone.

To show this, it will clearly be enough to show that on every
extremal ray of $\Lambda_n$ there exists a nonzero vector contained
in $\Lambda_n^*$. In other words, if we can construct a pair of
states that has a relative entropy vector on an extremal ray, for
all possible extremal rays of $\Lambda_n$, then due to approximate
dilutability we can find entropy vectors along all points of all
extremal rays. Since every point inside a cone can be made with a
positive linear combination of points from its extremal rays, we
obtain that every point inside the cone can be realised and
$\Lambda_n=\overline{\Lambda_n^*}$.

Achieving these states can be identified with classical secret
sharing schemes (see for example~\cite{Stinson}) as we will explain.
The formalism
for a secret sharing scheme can be defined as follows. Imagine a
defined secret bit that we want to share between a number of
participants. We want only certain so-called "authorised" groups
of participants to be able to recover the secret exactly, while
unauthorised groups of parties get no information about the
secret. It is clear that with every authorised group $\cal S$, any
group $\cal S' \supset \cal S$ will also be authorised. So, the
authorised groups will form an up-set called an $\emph{access
structure}$.

\begin{definition}
An $n$-party secret sharing scheme for a bit $b$ with access
structure $\emptyset \neq \mathbf{U} \subset 2^{[n]}, \emptyset
\notin \cal \mathbf{U}$, consists of the following
\begin{description}
  \item[(i)] \quad Random variables $X_1(b),X_2(b),X_3(b), \ldots ,X_n(b)$, each
        one associated with a participant labelled $1,\ldots, n$ in the
        secret sharing scheme. $X_i(b)$ takes values in a set ${\cal
        X}_i$.

  \item[(ii)] \quad For $\cal S \in \mathbf{U}$, denote $X^{\cal S}(b)
  = (X_i(b):i \in \cal S)$, the collection of shares accessible to the
  group $\cal S$

  \item[(iii)] \quad For each $\cal S \in \mathbf{U}$, there is a
  function $f_{\cal S}:{\cal X}^{\cal S} := \prod_{ i \in \cal S} {\cal
  X}_i \rightarrow \{0,1\}$ s.t. $f_{\cal S}(X^{\cal S}(b))=b$. For $\cal
  S \notin \mathbf{U}$ however, $X^{\cal S}(0)$ and $X^{\cal S}(1)$ have
  the same distributions.
\end{description}
\end{definition}

With this scheme the notion of an up-set is naturally included.
Since an authorised group of parties are allowed to recover the
secret, adding additional parties must also result in an authorised
group since the decoding function can be chosen only to act on the
previous authorised group. This is the defining feature of an
up-set. To relate this to a quantum information setting, we can
construct the following density matrix based on a secret sharing
scheme:
\begin{equation}
\label{eq:densmat} \rho(b) = \sum_{x_1\ldots x_n}
\Pr\{X_1(b)=x_1,\ldots,X_n(b)=x_n\}\ketbra{x_1}^1 \otimes
\ket\bra{x_2}^2 \otimes \cdots \otimes \ketbra{x_n}^n.
\end{equation}
The superscript on the terms of the tensor product denote the
label of the share. We denote a partial trace of the matrix as
\begin{equation}
\rho(b)^\mathcal{S}=\sum_{x_j : j \in {\cal S}} Pr \{X_j(b)=x_j,
\forall j \in {\cal S}\}\bigotimes_{j \in {\cal S}} \ketbra{x_j}{x_j}^j.
\end{equation}
$\rho(b)^{\cal S}$ has the following properties :
\begin{itemize}
 \item If ${\cal S} \in {\mathbf{U}}$ then the
supporting subspace of $\rho(0)^\mathcal{S}$ is orthogonal to that
of $\rho(1)^\mathcal{S}$ which allows the group ${\cal S}$ to
determine the secret bit exactly: $\rho(0)^{\cal S} \perp
\rho(1)^{\cal S}$. \item If $\mathcal{S} \neq \mathbf{U}$ then
$\rho(0)^\mathcal{S}=\rho(1)^\mathcal{S}$ and no information about
the secret can be achieved.
\end{itemize}
With this density matrix we can construct the following matrices
for use in relative entropy $D(\rho\|\sigma)$:
\begin{align}
  \label{eq:den2}
  \rho^{\cal S}   &= \rho(0)^{\cal S}, \\
  \label{eq:den3}
  \sigma^{\cal S} &= \frac{1}{2}\bigl(\rho(0)^{\cal S}+\rho(1)^{\cal S}\bigr).
\end{align}
Note that if $\mathcal{S}\notin \mathbf{U}$ then $\rho^{\cal
S}=\sigma^{\cal S}$ and the relative entropy is zero. For ${\cal
S} \in \mathbf{U}$, we can calculate the relative entropy as
follows:
\begin{equation}
D(\rho^{\cal S}\|\sigma^{\cal S})=\tr \bigg[\rho(0)^{\cal S} \log
\rho(0)^{\cal S} - \rho(0)^{\cal S} \log \bigg(
\frac{\rho(0)^{\cal S}}{2} + \frac{\rho(1)^{\cal S}}{2}
\bigg)\bigg].
\end{equation}
Using $\rho(0)^{\cal S} \perp \rho(1)^{\cal S}$.
\begin{equation}
D(\rho^{\cal S}\|\sigma^{\cal S})=\tr \bigg[\rho(0)^{\cal S} \log
\rho(0)^{\cal S} - \rho(0)^{\cal S} \log \frac{\rho(0)^{\cal
S}}{2} + \rho(0)^{\cal S} \log \frac{\rho(1)^{\cal S}}{2}\bigg].
\end{equation}
Since there are no elements in $\rho(0)^{\cal S}$ that are present
in $\rho(1)^{\cal S}$ the third term is zero. Hence expanding the
second term
\begin{align}
D(\rho^{\cal S}\|\sigma^{\cal S}) &= \tr \bigg[\rho(0)^{\cal S}
\log \rho(0)^{\cal S} - \rho(0)^{\cal S} \log
\rho(0)^{\cal S} + \rho(0)^{\cal S} (\log 2) \1 \bigg]\\
&= (\log 2) \tr [\rho(0)^{\cal S}]=1.
\end{align}

Note that the relative entropy is constant and independent of the
number of elements of ${\cal S}$. Hence we have states from which we
can produce relative entropies in the form of up-sets described in
Proposition 2 by simply realising a classical secret sharing scheme
with the required access structure. There exists a secret sharing
scheme for every up-set structure, in fact for every access
structure \cite{Shamir,ISN}. Therefore for each extremal ray of
$\Lambda_n$ there is a secret sharing scheme whose density operators
according to eqs.~(\ref{eq:densmat}), (\ref{eq:den2}) and
(\ref{eq:den3}) will produce the required relative entropy vector and
hence prove that each extremal ray is realisable. Hence we have
proved that $\overline{\Lambda_n^*}=\Lambda_n$ and thus that
monotonicity under restrictions is the only inequality satisfied by
relative entropies.

\section{Simple secret sharing: threshold schemes}
\label{sec:thres}
In this section we will describe a simple secret sharing scheme
for a specialised access structure known as a \emph{threshold}
scheme. We will then build upon this scheme showing how we can
construct schemes for any access structure. The threshold scheme
was discovered by Shamir \cite{Shamir} and allows parties to
recover a secret if and only if enough of the parties collaborate,
such that their number is beyond a predetermined threshold number
of parties. Each party is given a part of the secret which we call
a `share' of the secret. There is a total of $n$ shares, one share
for each party. A threshold value $k$ is also determined such that
if a number of parties get together and pool their shares, if the
number of shares they have are greater than or equal to $k$ then
they can recover the secret precisely. However, if the number of
shares is less than $k$, then no information can be extracted
about the secret. Accordingly, these schemed are called
$(n,k)$-threshold schemes, depending on the number of parties and the
desired threshold value. The construction of the threshold scheme
is outlined as follows. The premise for the scheme is based on
evaluations of a polynomial. Imagine the following polynomial.
\begin{equation}
y=a_0+a_1x+a_2x^2+a_3x^3+\ldots+a_{m-1}x^{m-1}
\end{equation}
We label $a_0$ as the secret value and the shares as evaluations
of this polynomial at different points. Geometry tells us that we
need exactly $m$ evaluations of this polynomial to determine the
coefficient $a_0$ and that if we have any fewer than $m$
evaluations any value of $a_0$ would fit the given points. This
means that if we have $m$ or more evaluations we know the secret
exactly and if we have fewer than $m$ evaluations we know nothing
about the secret. The evaluations of the polynomials becomes the
'shares' of the scheme and we perform $(n,k)$-threshold scheme the
calculations over a finite field. Here is a formulation of the
scheme extracted from the original paper by Shamir ~\cite{Shamir}.
\begin{itemize}
\item Choose a random $k-1$ degree polynomial
$y(x)=a_0+a_1x+a_2x^2+\ldots+a_{k-1}x^{k-1}$ and let $s$ be the secret
where $s=a_0$ i.e. $a_1,\ldots a_{k-1}$ are chosen independently
and uniformly from the field $GF(p)$ of $p$ elements (integer
modulo $p$)\item The shares are defined as $D_1=y(1),
D_2=y(2),\ldots,D_i=y(i),\ldots,D_n=y(n)$.

\item Any given subset
of $k$ of these $D_i$ values together with their indices can find
the coefficients of $y(x)$ by interpolation and hence find the
value of $s=y(0)$. \item Knowing $k-1$ or fewer shares will not
reveal what the value of $s$ as there exists polynomials that will
fit the given points in the polynomial and allow $a_0=0$ or
$a_0=1$ with every polynomial equally likely. 

\item We use a set
of integers modulo a prime number $p$ which forms a finite field
allowing interpolation. \item Given that the secret is an integer
we require $p$ to be larger than both max $s$ and $n$. \item If we
only have $k-1$ shares, there is one and only one polynomial that
can be constructed for each value of $s$ in $GF(p)$. Since each
polynomial is equally likely by construction, no information about
the secret can be gained.
\end{itemize}

This scheme can be easily translated to the quantum density matrix
defined in eq.~(\ref{eq:densmat}). Most of the probabilities in
the sum are zero except for the ones that are valid for a polynomial
fitting the secret value, with shares labeling that part of the sum.
This scheme has a very specific access structure, but we can expand
to more general access structures. Consider the number of parties
$p$, we can have $n>p$ so that we have more shares than parties,
allowing us to distribute multiple shares to single parties. This
allows us to have access structures not possible with the simple
access structure. Imagine that we require an access structure given
in Fig.~2. We require that B and C cannot recover the secret, however
if they pool their resources together they can. We also need A to be
able to recover the secret independently. Under the normal threshold
scheme, we need the threshold to be set at $k=1$ so that single
party A can recover the secret. However, this means B and C will
independently be also able to recover the secret so we cannot create
the required access structure.

However, if we use a scheme with more shares than parties, we can
achieve this access structure, see Example 5. Many up-sets can be
realised using this modified threshold scheme. The following
example provides the required threshold scheme and the resulting
density matrices.

\begin{example}
Imagine an $n=3$ system, each labelled by A,B and C respectively.
Consider also the following up-set representing an extremal ray.
This is Ray 6 as used in the previous section.
\begin{center}
\begin{tabular}[c]{|c|c|c|c|c|c|c|}
    \hline
    \ $v_A$ & $v_B$ & $v_C$ & $v_{AB}$ & $v_{AC}$ & $v_{BC}$ & $v_{ABC}$\\
    \hline
    \ 1    &   0  &   0  &  1   &   1   &   1   &    1\\
    \hline
\end{tabular}
\end{center}
\begin{figure}
\begin{center}
\includegraphics[scale=0.6,clip]{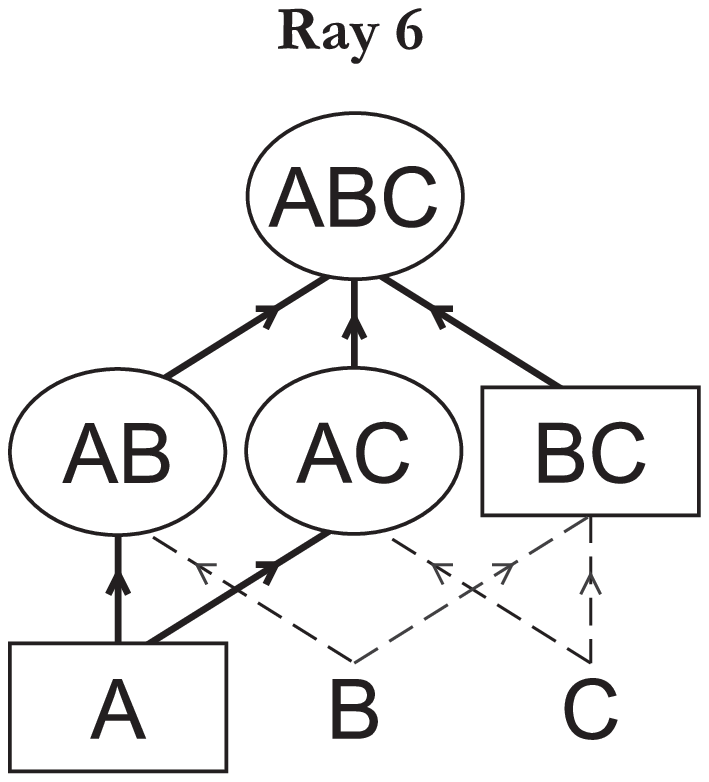}
\end{center}
\begin{caption}
{Diagram of up-set used in Example 5}
\end{caption}
\end{figure}
With this up-set we can now construct a secret sharing scheme to
represent it. One of the easiest constructions to understand is
the threshold scheme. The scheme required is a (4,2) threshold
scheme: 4 is the total number of shares, 2 shares or higher
required to construct secret. We distribute the shares as follows:
two shares to A and only one share to B and one to C. This leads
us to the required access structure as shown below.
\begin{center}
\begin{tabular}[c]{|c|c|c|c|c|c|c|c|}
    \hline
    \ &$A$ & $B$ & $C$ & $AB$ & $AC$ & $BC$ & $ABC$\\
    \hline
    \ Shares & 2    &   1  &   1  &  3   &   3   &   2   &    4\\
    \ Above threshold
    &$\checkmark$&$\times$&$\times$&$\checkmark$&$\checkmark$&$\checkmark$&$\checkmark$\\
    \hline
\end{tabular}
\end{center}
Since we have a total of four shares, we have to construct the
scheme of a finite field of 5. In this example calculations will
be assumed to be done over this finite field. Since the threshold
is two shares, we only need consider polynomials of order one,
since only two or more values are necessary to recover the
polynomial of order 1. Therefore the possible polynomials are as
follows.
\begin{align*}
y&=s\\
y&=s+x\\
y&=s+2x\\
y&=s+3x\\
y&=s+4x
\end{align*}
We can now embed this scheme into a quantum system. Each system
has the same number of qudits as the corresponding party has
shares, with $d$ being
large enough to incorporate the finite field values (i.e. in this
case d=5). For example system A has two qudits whereas
system B only has one. We now construct the density matrices
$\rho(0)$ and $\rho(1)$ as follows:
\begin{align}
  \rho(0) &= \frac{1}{5}\bigl(
                \ketbra{0000}{0000}+\ketbra{1234}{1234}+\ketbra{2413}{2413}
                                   +\ketbra{3142}{3142}+\ketbra{4321}{4321}\bigr) \\
  \rho(1) &= \frac{1}{5}\bigl(
                \ketbra{1111}{1111}+\ketbra{2340}{2340}+\ketbra{3024}{3024}
                                   +\ketbra{4203}{4203}+\ketbra{0432}{0432}\bigr)
\end{align}
$A$ has the first two qudits, $B$ the third and $C$ the fourth.
From this we can construct the overall system described
previously. We take $\rho=\rho(0)$ and
$\sigma=\frac{\rho(0)+\rho(1)}{2}$ as in eqs.~(\ref{eq:den2})
and (\ref{eq:den3}). As examples we may compute
\begin{align}
\rho_A  &=\frac{1}{5}\bigl(
           \ketbra{00}{00}+\ketbra{12}{12}+\ketbra{24}{24}+\ketbra{31}{31}+\ketbra{43}{43}\bigr),\\
\sigma_A&=\frac{1}{10}\bigl(
           \ketbra{00}{00}+\ketbra{12}{12}+\ketbra{24}{24}+\ketbra{31}{31}+\ketbra{43}{43}\bigr.
                                                                                        \nonumber\\
        &\phantom{===}\bigl.
          +\ketbra{11}{11}+\ketbra{23}{23}+\ketbra{30}{30}+\ketbra{42}{42}+\ketbra{04}{04}\bigr).
\end{align}
Therefore it can be verified that the relative entropy of party A
is $\log{2}$. Repeating this for party B.
\begin{equation}
\rho_B =
\frac{1}{5}\bigl(\ketbra{0}{0}+\ketbra{3}{3}+\ketbra{1}{1}+\ketbra{4}{4}
                                                          +\ketbra{2}{2}\bigr)=\sigma_B.
\end{equation}
Therefore the relative entropy for B is 0. All other relative
entropies can be verified in this way.
\end{example}
Thus giving unequal number of shares to the parties can achieve
more complicated access structures. However not all access
structures can be produced in this way. For example imagine that
we have a 4 parties A,B,C and D with number of shares in each
party being $a,b,c$ and $d$ respectively. We require that A and B
can recover the secret and that C and D can recover the secret but
no other two party combination. If A and B can recover the secret
then their combine total of shares must be greater than $k$ i.e.
$a+b\geq k$. Therefore either $a\geq \frac{k}{2}$ or $b\geq
\frac{k}{2}$. Similarly we can claim that $c\geq \frac{k}{2}$ or
$d\geq \frac{k}{2}$. Say that in this case $a\geq \frac{k}{2},
c\geq \frac{k}{2}$. Hence there exists another two party
combination, A and C, that have a number of shares greater than
$k$ and can recover the secret i.e. $a+c\geq k$. Therefore the
access structure is impossible to produce with this scheme.
However there are general methods for dealing with arbitrary
access structures~\cite{ISN,BenLei}. These allow us to represent any
extremal ray. One strategy is to create a hierarchy of threshold
schemes. Here we illustrate the strategy with an example.
\begin{example}
Imagine an $n=4$ system which we label $A,B,C$ and $D$ respectively.
Consider also the following up-set representing an extremal ray.
\begin{center}
\begin{tabular}[c]{|c|c|c|c|c|c|c|c|c|c|c|c|c|c|c|}
    \hline
    \ $v_A$ & $v_B$ & $v_C$ & $v_D$ & $v_{AB}$ & $v_{AC}$ & $v_{AD}$ & $v_{BC}$&$v_{BD}$&$v_{CD}$&$v_{ABC}$&$v_{ABD}$&$v_{ACD}$&$v_{BCD}$&$v_{ABCD}$\\
    \hline
    \ 0     &   0   &   0   &  0   &   1   &   0   &    0&0&0&1&1&1&1&1&1\\
    \hline
\end{tabular}
\end{center}
Note that access structure representing this ray requires that no
single party has access to the secret and only parties A and B
collaborating, and C and D collaborating will be authorised. Also
any greater number of parties will always contain an authorised
group and are therefore also authorised. The required access
structure can be represented by two schemes. This in illustrated
in Fig.~3.
\begin{figure}
\begin{center}
\includegraphics[scale=0.34,clip]{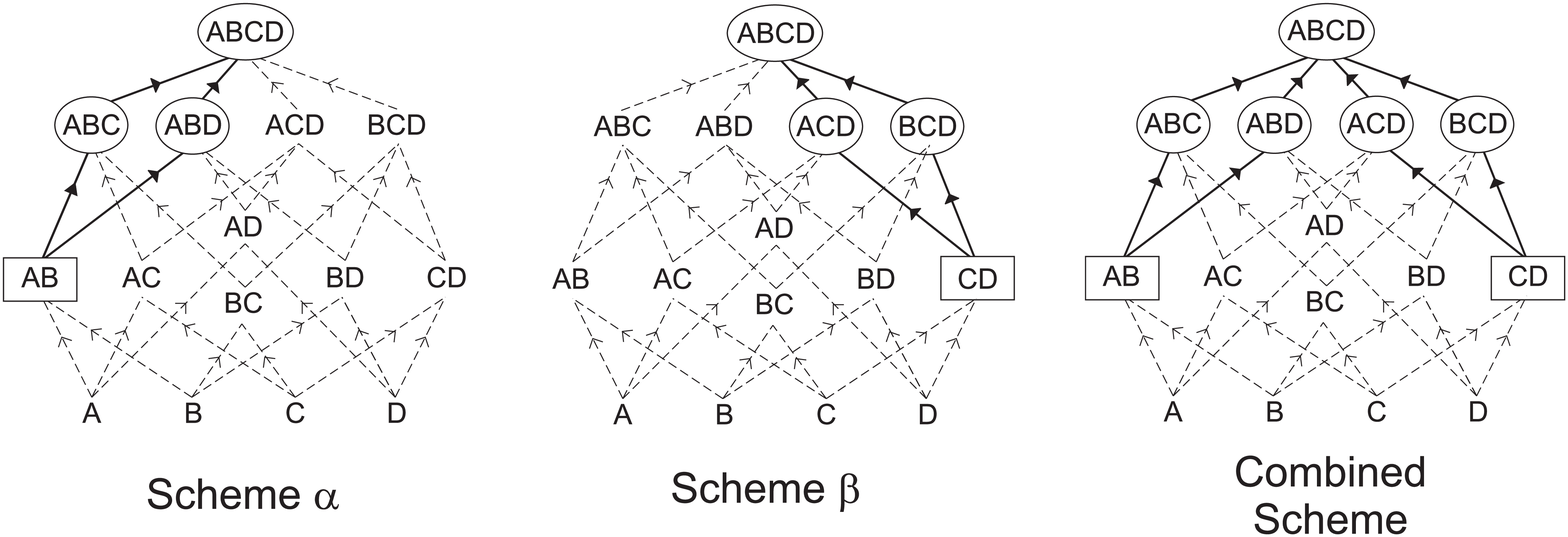}
\end{center}
\begin{caption}
{Diagram of up-set used in Example 6}
\end{caption}
\end{figure}
Each scheme requires a $(2,2)$-threshold scheme, 2 total number of
shares with a threshold for recovering the secret of 2 shares. We
distribute the shares as follows : in one scheme (scheme $\alpha$)
we give 1 share to A and 1 share to B. In the other scheme (scheme
$\beta$) we give 1 share to C and 1 share to D. This ensures that
the secret can be recovered by authorised parties via at least one
of the schemes reaching threshold, shown below.
\begin{center}
\begin{tabular}[c]{|c|c|c|c|c|c|c|c|c|c|c|c|c|c|c|c|}
    \hline
    \ &       $A$ & $B$ & $C$ & $D$ & $AB$ & $AC$ & $AD$ & $BC$ & $BD$ & $CD$ & $ABC$ & $ABD$ & $ACD$ & $BCD$ & $ABCD$\\
    \hline
    \ Shares(scheme $\alpha$) & 1  &  1  &  0  &  0  &  2   &  1   &  1   &  1   &  1   &  0   &   2   &   2   &   1   &   1   &   2   \\
    \ Shares(scheme $\beta$) & 0  &  0  &  1  &  1  &  0   &  1   &  1   &  1   &  1   &  2   &   1   &   1   &   2   &   2   &   2   \\
    \ Above threshold
    &$\times$&$\times$&$\times$&$\times$&$\checkmark_\alpha$&$\times$&$\times$&$\times$&$\times$&$\checkmark_\beta$&$\checkmark_\alpha$&$\checkmark_\alpha$&$\checkmark_\beta$&$\checkmark_\beta$&$\checkmark_{\alpha,\beta}$\\
    \hline
\end{tabular}
\end{center}
Since we have a total of two shares for each scheme, we construct
the scheme using a finite field of 3 elements. From now on
calculations will be assumed to be done over this finite field.
Since the threshold is two shares, we only need consider
polynomials of order one, since only two or more coordinates are
necessary to recover the polynomial of order 1. Therefore the
possible polynomials are as follows.
\begin{align*}
y&=s\\
y&=s+x\\
y&=s+2x
\end{align*}
In the construction of the quantum density matrix we need to
consider all possible set of shares the individual parties can
have. The possible combinations are presented in the following
tables.
\begin{table}[!h]
\begin{minipage}{3in}
\begin{tabular}{|c|c|cccc|}
\hline
\ Scheme $ \alpha$&Scheme $\beta$ &$A$&$B$&$C$&$D$\\
\hline
\ $y=s$&$y=s$       &$0\ast$&$0\ast$&$\ast0$&$\ast0$\\
\ $y=s$&$y=s+x$     &$0\ast$&$0\ast$&$\ast1$&$\ast2$\\
\ $y=s$&$y=s+2x$    &$0\ast$&$0\ast$&$\ast2$&$\ast1$\\
\ $y=s+x$&$y=s$     &$1\ast$&$2\ast$&$\ast0$&$\ast0$\\
\ $y=s+x$&$y=s+x$   &$1\ast$&$2\ast$&$\ast1$&$\ast2$\\
\ $y=s+x$&$y=s+2x$  &$1\ast$&$2\ast$&$\ast2$&$\ast1$\\
\ $y=s+2x$&$y=s$    &$2\ast$&$1\ast$&$\ast0$&$\ast0$\\
\ $y=s+2x$&$y=s+x$  &$2\ast$&$1\ast$&$\ast1$&$\ast2$\\
\ $y=s+2x$&$y=s+2x$ &$2\ast$&$1\ast$&$\ast2$&$\ast1$\\
\hline
\end{tabular}
\begin{caption}
{All possible shares for $s=0$}
\end{caption}
\end{minipage}
\begin{minipage}{3in}
\begin{tabular}{|c|c|cccc|}
\hline
\ Scheme $ \alpha$&Scheme $\beta$ &$A$&$B$&$C$&$D$\\
\hline
\ $y=s$&$y=s$      &$1\ast$&$1\ast$&$\ast1$&$\ast1$\\
\ $y=s$&$y=s+x$    &$1\ast$&$1\ast$&$\ast2$&$\ast0$\\
\ $y=s$&$y=s+2x$   &$1\ast$&$1\ast$&$\ast0$&$\ast2$\\
\ $y=s+x$&$y=s$    &$2\ast$&$0\ast$&$\ast1$&$\ast1$\\
\ $y=s+x$&$y=s+x$  &$2\ast$&$0\ast$&$\ast2$&$\ast0$\\
\ $y=s+x$&$y=s+2x$ &$2\ast$&$0\ast$&$\ast0$&$\ast2$\\
\ $y=s+2x$&$y=s$   &$0\ast$&$2\ast$&$\ast1$&$\ast1$\\
\ $y=s+2x$&$y=s+x$ &$0\ast$&$2\ast$&$\ast2$&$\ast0$\\
\ $y=s+2x$&$y=s+2x$&$0\ast$&$2\ast$&$\ast0$&$\ast2$\\
\hline
\end{tabular}
\begin{caption}
{All possible shares for $s=1$}
\end{caption}
\end{minipage}
\end{table}

Each party has two registers, one for each scheme. If a party has no
share then the register associated with that scheme is put into a
fixed state (here
$\ket{\ast}_A,\ket{\ast}_B,\ket{\ast}_C,\ket{\ast}_D$) which is
uncorrelated to the variables for that scheme. Thus the density
matrix $\rho(0)$ is

\begin{align*}
\rho(0)=\frac{1}{9}&
\bigl( \ \proj{0{\ast}}_A\,\proj{0{\ast}}_B\,\proj{{\ast}0}_C\,\proj{{\ast}0}_D
         +\proj{0{\ast}}_A\,\proj{0{\ast}}_B\,\proj{{\ast}1}_C\,\proj{{\ast}2}_D \\
      &
       +\proj{0{\ast}}_A\,\proj{0{\ast}}_B\,\proj{{\ast}2}_C\,\proj{{\ast}1}_D
         +\proj{1{\ast}}_A\,\proj{2{\ast}}_B\,\proj{{\ast}0}_C\,\proj{{\ast}0}_D \\
      &
       +\proj{1{\ast}}_A\,\proj{2{\ast}}_B\,\proj{{\ast}1}_C\,\proj{{\ast}2}_D
         +\proj{1{\ast}}_A\,\proj{2{\ast}}_B\,\proj{{\ast}2}_C\,\proj{{\ast}1}_D \\
      &
       +\proj{2{\ast}}_A\,\proj{1{\ast}}_B\,\proj{{\ast}0}_C\,\proj{{\ast}0}_D
         +\proj{2{\ast}}_A\,\proj{1{\ast}}_B\,\proj{{\ast}1}_C\,\proj{{\ast}2}_D \\
      &
       \phantom{+\proj{0{\ast}}_A\,\proj{2{\ast}}_B\,\proj{{\ast}0}_C\,\proj{{\ast}2}_D}
         +\proj{2{\ast}}_A\,\proj{1{\ast}}_B\,\proj{{\ast}2}_C\,\proj{{\ast}1}_D \bigr).
\end{align*}

Similarly we can construct $\rho(1)$ by repeating the process but
setting the secret bit to be 1, i.e.
$\ket{1{\ast}}_A \ket{1{\ast}}_B \ket{{\ast}1}_C \ket{{\ast}1}_D$
etc., leading to the density matrix $\rho(1)$:

\begin{align*}
\rho(1)=\frac{1}{9}&
\bigl( \ \proj{1{\ast}}_A\,\proj{1{\ast}}_B\,\proj{{\ast}1}_C\,\proj{{\ast}1}_D 
         +\proj{1{\ast}}_A\,\proj{1{\ast}}_B\,\proj{{\ast}2}_C\,\proj{{\ast}0}_D \\
      &
       +\proj{1{\ast}}_A\,\proj{1{\ast}}_B\,\proj{{\ast}0}_C\,\proj{{\ast}2}_D
         +\proj{2{\ast}}_A\,\proj{0{\ast}}_B\,\proj{{\ast}1}_C\,\proj{{\ast}1}_D \\
      &
       +\proj{2{\ast}}_A\,\proj{0{\ast}}_B\,\proj{{\ast}2}_C\,\proj{{\ast}0}_D
         +\proj{2{\ast}}_A\,\proj{0{\ast}}_B\,\proj{{\ast}0}_C\,\proj{{\ast}2}_D \\
      &
       +\proj{0{\ast}}_A\,\proj{2{\ast}}_B\,\proj{{\ast}1}_C\,\proj{{\ast}1}_D
         +\proj{0{\ast}}_A\,\proj{2{\ast}}_B\,\proj{{\ast}2}_C\,\proj{{\ast}0}_D \\
      &
       \phantom{+\proj{0{\ast}}_A\,\proj{2{\ast}}_B\,\proj{{\ast}0}_C\,\proj{{\ast}2}_D}
         +\proj{0{\ast}}_A\,\proj{2{\ast}}_B\,\proj{{\ast}0}_C\,\proj{{\ast}2}_D \bigr).
\end{align*}

We notice that in both states $\rho(0)$ and $\rho(1)$ in this
example the state $\ket{\ast}_A\ket{\ast}_B\ket{\ast}_C\ket{\ast}_D$
factors out so that we could equally well take

\begin{align}
\rho(0) &= \frac{1}{9}\bigl(
            \ketbra{0000}{0000}+\ketbra{0012}{0012}+\ketbra{0021}{0021}
                                                   +\ketbra{1200}{1200} \nonumber  \\
        &\phantom{==}
           +\ketbra{1212}{1212}+\ketbra{1221}{1221}
               +\ketbra{2100}{2100}+\ketbra{2112}{2112}+\ketbra{2121}{2121} \bigr), \\
\rho(1) &= \frac{1}{9}\bigl(
            \ketbra{1111}{1111}+\ketbra{1120}{1120}+\ketbra{1102}{1102}
                                                   +\ketbra{2011}{2011} \nonumber  \\
        &\phantom{==}
            +\ketbra{2020}{2020}+\ketbra{2002}{2002}
               +\ketbra{0211}{0211}+\ketbra{0220}{0220}+\ketbra{0202}{0202} \bigr).
\end{align}

[Note however that in more complicated examples parties need shares from
more than one scheme.] From this we can construct the overall system
described previously, and for example for parties $AB$.
\begin{align}
  \rho_{AB}   &= \frac{1}{3}\bigl(
                   \ketbra{00}{00}+\ketbra{12}{12}+\ketbra{21}{21}\bigr), \\
  \sigma_{AB} &= \frac{1}{6}\bigl(
                   \ketbra{00}{00}+\ketbra{12}{12}+\ketbra{21}{21}+
                    \ketbra{11}{11}+\ketbra{20}{20}+\ketbra{02}{02} \bigr).
\end{align}
Therefore it can be verified that the relative entropy of parties AB
is $\log{2}$. Repeating this for parties BC,
\begin{equation}
\rho_{BC} =
\frac{1}{6}\bigl(\ketbra{11}{11}+\ketbra{12}{12}+\ketbra{10}{10}+\ketbra{02}{02}
+\ketbra{00}{00}+\ketbra{21}{21}+\ketbra{22}{22}+\ketbra{20}{20}\bigr)=\sigma_{BC}.
\end{equation}
Therefore the relative entropy for BC is 0. All other relative
entropies can be verified in this way.
\end{example}

The idea of using a hierarchy of threshold schemes was discovered by
Ito, Saito and Nishizeki \cite{ISN} and requires an exponential
number of threshold schemes to represent an access structure. This
number of schemes required is irrelevant as long as a scheme exists
and we can create the corresponding density matrix. A simpler
general access structure was found by Benaloh and Leichter
\cite{BenLei}, which does not use threshold schemes but can be
directly translated to the required density matrices in
eq.~(\ref{eq:densmat}).

\section{Conclusion}
\label{sec:coda}
In this paper, we have determined the set of all
relative entropy vectors for general states on (general) $n$-party
systems: it coincides with the convex cone defined by
non-negativity and monotonicity of the relative entropy. We have
done this by first showing that the former set in is indeed a
convex cone, and then demonstrating that every extremal ray in the
latter cone is realised by a specific pair of states. These
extremal rays are characterised by up-sets in $2^{[n]}$, and the
pairs of states correspond to (classical) secret sharing schemes.

A particular consequence is that the cone of relative entropy
vectors is the same for quantum states and for classical
probability distributions. This is in marked contrast to the case
of entropy vectors, where even for $n=2$ classical and quantum
entropy cone differ~\cite{Pippenger}.

Beyond the characterisation in terms of convex geometry, our
result also means that, apart from monotonicity, there can be no
other univeral relation between the relative entropy values of the
reduced states in a composite systems (except that is follows
trivially from monotonicty). In this sense, quantum and classical
relative entropy is completely characterised by the monotonicity
relation.

We are now in a position to go back to our assumption of finite
dimensional systems and the demand that all relative entropies are
finite. Clearly, if some of the parties are described by infinite
dimensional quantum systems, we still have
monotonicity~\cite{Uhlmann}, so the relative entropy vectors are all
within the Lindblad-Uhlmann cone. In this case, and even in the
finite dimensional case some entries in a relative entropy vector
may be positive infinity. However, even this does not present a
problem, once we realise that the groups where the value is infinite
form an up-set, so the vector can indeed be obtained as a limit of
finite relative entropy vectors in the Lindblad-Uhlmann cone.

Another mathematical peculiarity is the following: From the proof
of achievability of all extremal ray of the Lindblad-Uhlmann
cone, we discover that every
point in the entropy cone is achievable rather than infinitely
approximated, i.e. $\Lambda_n = \Lambda_n^*$. This is due to the fact that
every point on all exremal rays can be attained. To see, this,
simply choose $\rho(0)$ and $\rho(1)$ in eq.~(\ref{eq:den3})
with different weights $p$ and $1-p$ ($0\leq p\leq 1$). Then the
calculation following that equation shows that
the relative entropy is either $H_2(p)$ or $0$ depending
on whether ${\cal S}$ is an authorised set or not. By additivity
in Lemma~\ref{lemma:cone} we obtain that every point on the extremal
rays is realised, hence every point in the Lindblad-Uhlmann cone.

\par\medskip
We conclude the paper by commenting briefly on possible connections
of our result to the entropy cone, and possibly to the relative
entropy of entanglement. In the above arguments we have often used
the formula $D(\rho\|\sigma) = -S(\rho)-\tr\rho\log\sigma$, which
means that if we make the restriction $\sigma = \frac{1}{d}\1$, the
maximally mixed state in $d$ dimensions, the relative entropies (now
dependent only on $\rho$) evaluate to $\log d - S(\rho)$. Going
through the proof of Lemma~\ref{lemma:cone} we see that for any number $n$ of
parties, the set of all these relative entropy vectors is also a
convex cone, and one might think that its relations would capture
all inequalities for the entropy. That this is too optimistic a
hope, is indicated by the fact that the relative entropy is
expressed by the entropy and a term beyond what can be expressed by
general entropies alone (essentially the log of the rank). And it is
indeed not the case, since for example the nonegativity of the
relative entropy translates into $S(\rho) \leq \log d$. However, the
fundamental fact that the entropy $S(\rho)$ is nonnegative, is not
captured at all, since that would require an upper bound on the
relative entropy depending on the dimension. Still, there may be
some less stringent relation between the entropy and the relative
entropy cones, whose existence we would like to advertise as an open
problem.

\acknowledgments
BI was supported by the U.K. Engineering and
Physical Sciences Research Council. NL and AW acknowledge support by
the EU project RESQ and the U.K.~EPSRC's IRC QIP.

\end{document}